# Frozen Superparaelectric State of Local Polar Regions in $GdMn_2O_5$ and $Gd_{0.8}Ce_{0.2}Mn_2O_5$

V. A. Sanina*, B. Kh. Khannanov, E. I. Golovenchits, and M. P. Shcheglov

*Ioffe Physico-Technical Institute, Russian Academy of Sciences, ul. Politekhnicheskaya 26, St. Petersburg, 194021 Russia*

*e-mail: sanina@mail.ioffe.ru

A comparative study of the dielectric properties and electric polarization of multiferroics $GdMn_2O_5$ and $Gd_{0.8}Ce_{0.2}Mn_2O_5$ has been carried out in the temperature range 5–330 K. The polarization properties in the ferroelectric state that forms due to a charge ordering and exchange striction have been studied at $T \leq T_C = 30$ K. The properties of the restricted polar phase separation domains formed in the crystals containing ions $Mn^{3+}$ and $Mn^{4+}$ have been studied, too. These domains exhibit the electric polarization in the temperature range from 5 K to some temperatures $T_f \gg T_C$. Such a high-temperature polarization is due to the frozen superparaelectric state of the restricted polar domains.

## 1. INTRODUCTION

Manganites $RMn_2O_5$ ($R$ is a rare-earth ion) are typical representatives of multiferroics in which the ferroelectric ordering is induced and is controlled by a magnetic order. The characteristic values of the Curie $T_C$ and Neel $T_N$ temperatures are 30–35 K and 40–45 K, respectively [1, 2]. Up to now, it was assumed that $RMn_2O_5$ is characterized at room temperature by centrosymmetric sp. gr. *Pbam* that forbids the existence of electric polarization. To explain the polarization observed in $RMn_2O_5$ at $T < 30$–35 K, the model of exchange striction caused by charge ordering of manganese ions with different valences ($Mn^{3+}$ and $Mn^{4+}$) along the axis $b$ was developed [3]. Thus, it was assumed that the electric polarization in $RMn_2O_5$ exists only at temperatures $T \leq T_C$, at which specific long-range magnetic order and the charge ordering induce the lattice noncentrosymmetry.

A specific feature of $RMn_2O_5$ is the existence of the same numbers of manganese ions $Mn^{3+}$ (containing $3t_{2g}$, $1e_g$ electrons on the $3d$ shell) and $Mn^{4+}$ ($3t_{2g}$, $0e_g$ electrons), which provides the conditions for the formation of a dielectric charge ordering. Ions $Mn^{4+}$ have the octahedral oxygen environment and are located in layers with $z = 0.25c$ and $(1 - z) = 0.75c$. Ions $Mn^{3+}$ have the off-center local environment as pentagon pyramids and are located in layers with $z = 0.5c$. Ions $R^{3+}$ with the environment similar to that of Mn3+ are located in layers with $z = 0$ [4]. The charge ordering and a finite probability of transferring $e_g$ electrons between $Mn^{3+}$–$Mn^{4+}$ ion pairs (double exchange [5,6]) are key factors that determine electric polar states in $R$Mn$_2$O$_5$ at all temperatures. As was noted above, the low-temperature ferroelectric state at $T \leq T_C$ was due to the charge ordering along axis $b$ [3]. On the other hand, the transfer of $e_g$ electrons between $Mn^{3+}$–$Mn^{4+}$ ion pairs disposed in neighboring layers perpendicular to the axis $c$ leads to the formation of restricted polar phase separation domains with a different distribution of ions $Mn^{3+}$–$Mn^{4+}$ as compared to the initial crystal matrix. These local domains in $R$Mn$_2$O$_5$ are polar and exist from the lowest temperatures to temperatures higher than room temperature [7–10].

In [11], a series of $R$Mn$_2$O$_5$ crystals with different $R$ ions was studied by resonance synchrotron X-ray diffraction and geometric optimization based on first principal calculations at room temperature. Those authors observed the intense reflections corresponding to sp. gr. *Pbam* and significantly weaker reflections that could not be described by the central symmetry, and they concluded, using physical arguments, that the real symmetry of $R$Mn$_2$O$_5$ was described by the monoclinic noncentrosymmetric sp.gr. *Pm* that allows the existence of the polarization along axis $b$. This meant that, up to room temperature, the paramagnetic phase of $R$Mn$_2$O$_5$ must also have a ferroelectric ordering of different nature that, at $T <$ 30–35 K, coexisted with a more intense electric polarization of the exchange–striction nature. The nature of this additional ordering was not discussed in [11].

In [7–10], the temperature evolution of the dielectric properties (the dielectric permittivity and the conductivity) of a number of $R$Mn$_2$O$_5$ ($R$ = Eu, Bi, and Gd) were studied in the wide temperature range 5–330 K. Two facts have been established, as follows.

First, the free-dispersion maxima of the dielectric permittivity (ε') and the dielectric losses (ε'') characteristic of the phase transition to the ordered ferroelectric state were observed along the axis $b$ only near $T = T_C \approx$ 30–35 K. Such maxima were not observed along all the axes in the temperature range 35–330 K. This demonstrated that, in this temperature range, another high-temperature phase transition to the ordered ferroelectric state did not take place.

Second, the frequency-dependent anomalies of the dielectric permittivity and the conductivity characteristic of the restricted polar domains were observed in the paraelectric region in the temperature range 100–330 K. The domain sizes were sufficient that a structural ordering different than that in the matrix appeared. This was demonstrated by the splitting of the Bragg peaks into two reflections [7–10]. These results made it possible to argue that the superparaelectric state was detected in the paraelectric region of $R$Mn$_2$O$_5$ ($R$ = Eu, Bi, and Gd).

The restricted polar domains form in $R$Mn$_2$O$_5$ containing ions Mn$^{3+}$ and Mn$^{4+}$ due to the phase separation processes similar to those in manganites La$A$MnO$_3$ ($A$ = Sr, Ca, and Ba) [6, 12]. The dynamically equilibrated restricted phase separation domains form in the initial crystal matrix spontaneously due to self-organization. Doping of $R$Mn$_2$O$_5$ ($R$ = Eu, Gd) with Ce$^{4+}$ ions increases the concentration of such domains, and, at temperatures higher than 180 K, the interaction appears between restricted polar domains isolated before and forms the 2D superstructures perpendicular to axis $c$. In these superstructures, the initial matrix layers and the phase separation domains alternated. The layer widths were 700–900 Å at room temperature [7, 8]. At low temperatures ($T \leq T_C \approx 30$–40 K), the restricted phase separation domains were isolated 1D superlattices of ferromagnetic layers containing ions Mn$^{3+}$ and Mn$^{4+}$ in various proportions. In the superlattices, a set of ferromagnetic resonances and the electric polarization [9, 10, 13–15] were observed in the direction of the magnetic and electric fields, respectively.

In [9, 10], it was shown that, in $R$Mn$_2$O$_5$, a frozen superparaelectric state appeared below some temperatures different from various crystal axes in the paramagnetic temperature range. The response of this state to applied electric field $E$ has the shape of the hysteresis loops of the electric polarization oriented along the field. The frozen superparaelectric state of the restricted ferroelectric domains in a dielectric centrosymmetric matrix was considered theoretically in [16], but it was observed experimentally for the first time in $R$Mn$_2$O$_5$ ($R$ = Gd, Bi) [9, 10]. The authors of [9, 10] developed the PUND (Positive Up Negative Down) method of measuring hysteresis loops adapted for the case of studying the electric polarization due to the existence of restricted dynamical polar domains. This method allows us to subtract the contribution of the conductivity of these domains from the hysteresis loops.

This paper presents the results of the comparative study of the dielectric properties and the electric polarization of GdMn$_2$O$_5$ (GMO) and Gd$_{0.8}$Ce$_{0.2}$Mn$_2$O$_5$ (GCMO) in the temperature range 5–330 K. We present the results of analyzing the electric polarization measured by two different methods: the thermoactivated pyrocurrent and the hysteresis loop (PUND) methods. This comparison enabled us to separate the contributions to the polarization of the low-temperature polar order of the exchange-striction nature and the frozen superparaelectric states of the restricted polar domains. This also made it possible to understand the properties of the polarization of the restricted polar domains formed in different ways of applying an external electric field in two these measurement methods.

The choice of the materials for the study was due to the following circumstances. GMO demonstrated anomalous high polarization along axis b ($P_b$ = 0.25–0.35 μC/cm$^2$) at $T \leq 30$ K [8, 17, 18]. The usual values of $P_b$ in $R$Mn$_2$O$_5$ are 0.03–0.05 μC/cm$^2$ [2]. GMO also has unusual magnetic properties, as compared to other $R$Mn$_2$O$_5$. In [17], it was shown that both its magnetic subsystems (Mn and Gd) strongly interacted to one another, forming the general order parameter. A noncommensurated magnetic structure with the wave vector $q$ = (0.49, 0, 0.18) forms in the temperature range $T_{N1}$ = 33 K – $T_{N2}$ = 30 K. At $T = T_{N2}$, the lock-in transition to the commensurate magnetic phase with wave vector $q$ = (1/2, 0, 0) that exists to the lowest temperatures occurs. The polar order in GMO is observed at $T \leq T_C = T_{N2}$ [17]. Note that the spectrum of antiferromagnetic resonance characteristic of a uniaxial collinear antiferromagnetic structure with wave vector q=(1/2, 0, 0) was observed in GMO at $T \leq 30$ K [19,20].

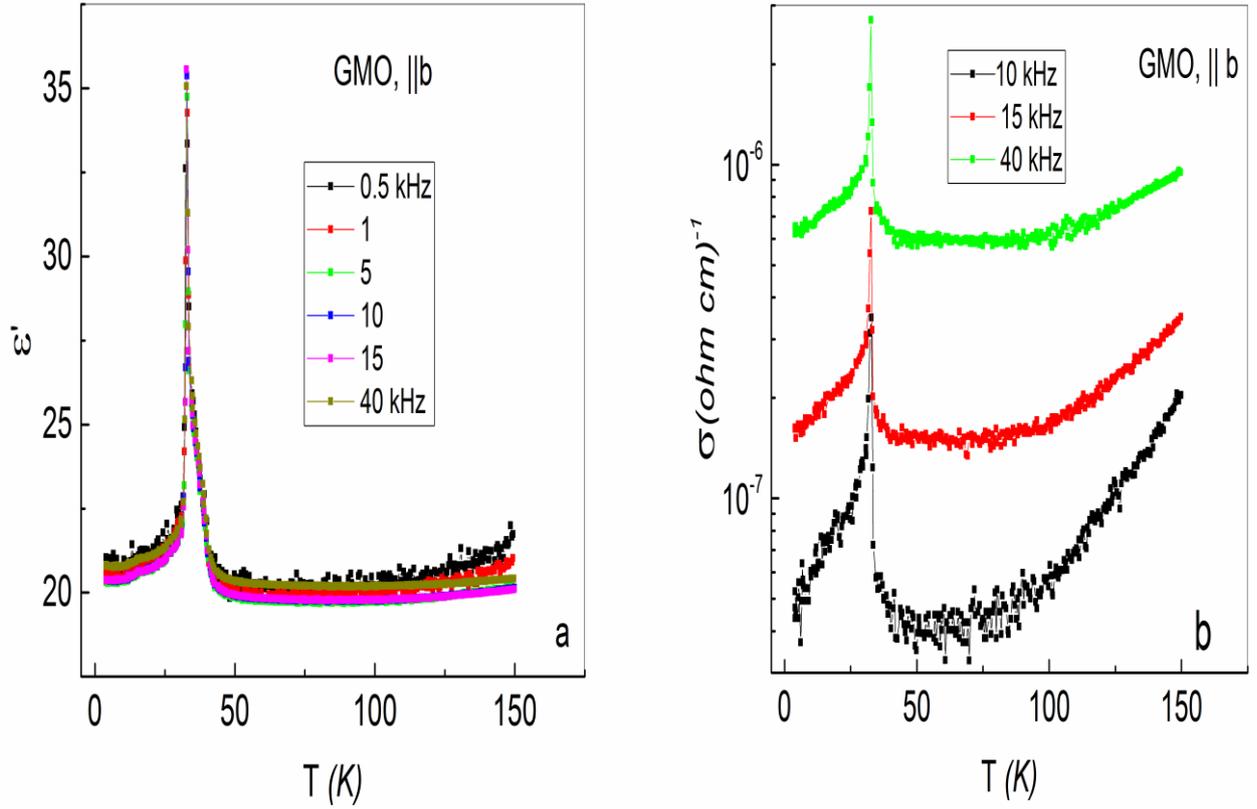

Fig. 1. Temperature dependences of (a) dielectric permittivity $\varepsilon'$ and (b) conductivity $\sigma$ for GMO measured along axis *b* at the frequencies indicated in the plots.

Unusual magnetic and polar states in GMO are due to the properties of the ground states of ions $Gd^{3+}$ ($^8S_{7/2}$) with large spin $S = 7/2$. These ions weakly interact with the lattice, but cause the strong uniform Gd–Mn exchange. This exchange increases the exchange striction and the polar order along the axis *b*. The comparative analysis of the magnetic properties of GMO and GCMO was performed in [15]. The magnetization of the doped crystal was slightly lower, but it, as before, is determined substantially by the magnetic Gd ions. The dilution of ions $Gd^{3+}$ with $Ce^{4+}$ ions leads to the disappearance of ordering in the Gd subsystem near 13 K. The Curie–Weiss temperature $T_{CW}$ almost coincides with $T_N$ of GMO; i.e., the magnetic states of both the crystals are not frustrated.

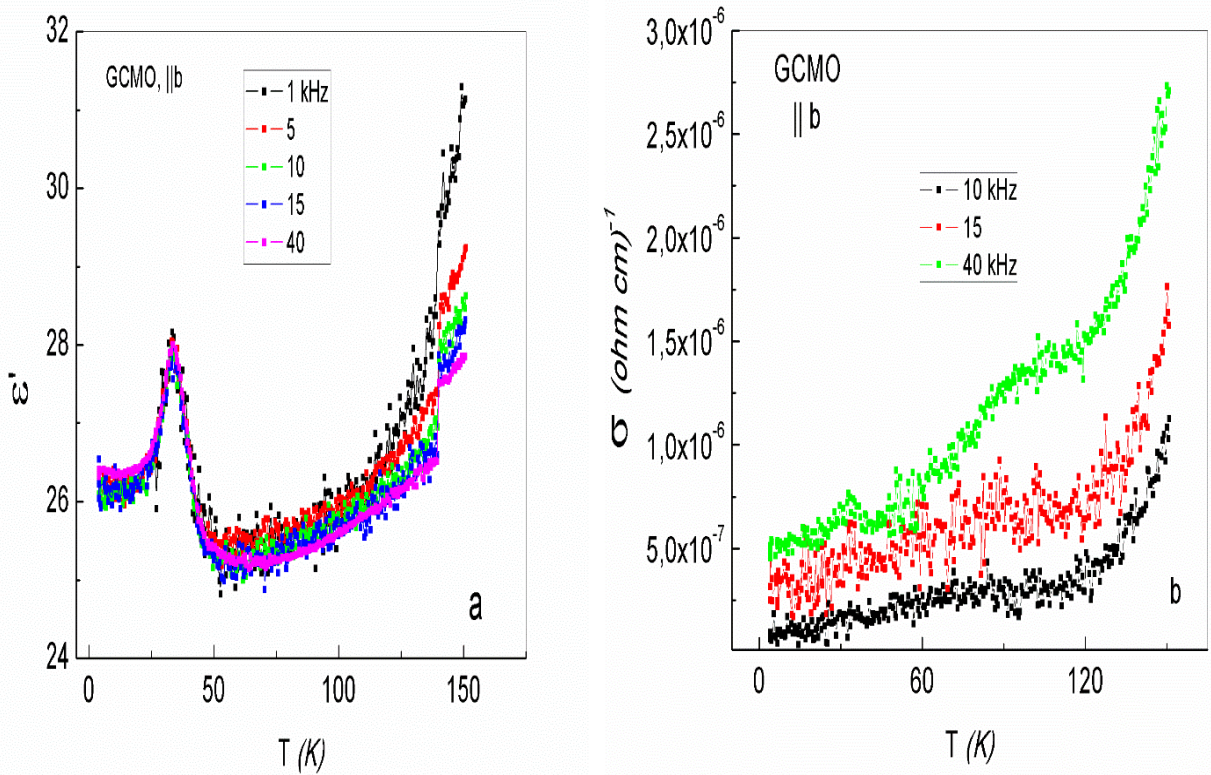

Fig. 2. Temperature dependences of (a) dielectric permittivity ε' and (b) conductivity σ for GCMO measured along axis b at the frequencies indicated in the plots.

It was interesting to carry out a comparative study of the electric polarization induced by the restricted polar phase separation domain in GMO and GCMO in which the domain concentrations are different.

2. EXPERIMENTAL RESULTS AND THEIR ANALYSIS

The GMO and GCMO single crystals were grown by spontaneous crystallization [21, 22]. They were 2–3-mm-thick plates with area of 3–5 mm$^2$. To measure the dielectric properties and the polarization, we fabricated 0.3–0.6-mm-thick flat capacitors with area of 3–4 mm2. The dielectric properties and the polarization were measured using a Good Will LCR-819 impedance meter in the frequency range 0.5–50 kHz and in the temperature range 5–330 K. The electric polarization was measured by two methods: the thermoactivated pyrocurrent and so-called PUND methods [23–25]. In the first case, the polarization was measured by a Keithly 6514 electrometer during heating of the sample at a constant rate of varying the temperature after preliminary cooling of the sample in a polarizing electric field. The polarization was determined by the integration of the temperature dependence of the pyrocurrent. In the second case, we used the PUND method adapted to the measurement of the polarization of the restricted polar domains with a local conductivity [9, 10].

2.1. DIELECTRIC PROPERTIES of GMO and GCMO

Firstly, we consider the data for axis *b* that is actual for GMO. Figures 1a and 1b show the temperature dependences of the dielectric permittivity $\varepsilon'$ and conductivity $\sigma$ for a number of frequencies. Figures 2a and 2b show the same dependences for GCMO at the same frequencies. A comparison of Figs. 1a and 2a shows that, near $T_C \approx 30$ K, both GMO and GCMO demonstrated free-dispersion maxima $\varepsilon'$ characteristic of the phase transition to the ferroelectric state. In this case, the maximum in GMO is significantly more intense and narrow as compared to that in GCMO. This means that the polar order in GMO at $T \leq T_C = 30$ K is significantly more homogeneous. It seems likely that the main sources of the inhomogeneity of the polar state in both the crystals are the phase separation domains the concentration of which in the doped crystal is significantly higher. The polar orders at $T \leq T_C$ in both the crystals are referred to their initial matrix. A noticeable increase in $\varepsilon'$ with temperature in both the crystals starts at $T > 125$ K. Figures 1b and 2b show the temperature dependences of the conductivity for GMO and GCMO, respectively.

We are dealing with the real part of conductivity $\sigma_1 = \omega\varepsilon''\varepsilon_0$ [26] that is calculated from the dielectric losses $\varepsilon''$ (we measured the dielectric loss tangent $tg\delta = \varepsilon''/\varepsilon'$). Here, $\omega$ is the angular frequency and $\varepsilon_0$ is the dielectric permittivity of free space. Conductivity $\sigma_1$, denoted further as $\sigma$, is dependent on the frequency and the temperature. The low-frequency part of the conductivity is free-dispersion and is referred to the percolation conductivity $\sigma_{dc}$. Conductivity $\sigma_{ac}$ has a frequency dispersion. In our case (Figs. 1b and 2b), the higher the frequency, the higher the conductivity. This frequency dispersion is characteristic of a local conductivity in restricted crystal domains with an energy barrier at their boundaries [26]. We assume that, in our case, the local conductivity is due to the phase separation domains, and the percolation conductivity is referred to the initial crystal matrix. Near 30 K, GMO demonstrates the free-dispersion maximum of conductivity (Fig. 1b) that is due to the maximum of $\varepsilon''$ near the ferroelectric transition; in GCMO, this maximum is hardly visible. The background conductivity that is thought to be related to the phase separation domains in GCMO is significantly higher than that in GMO (Figs. 1b and 2b). A marked increase in the conductivity with temperature in both the crystals starts at T> 125 K.

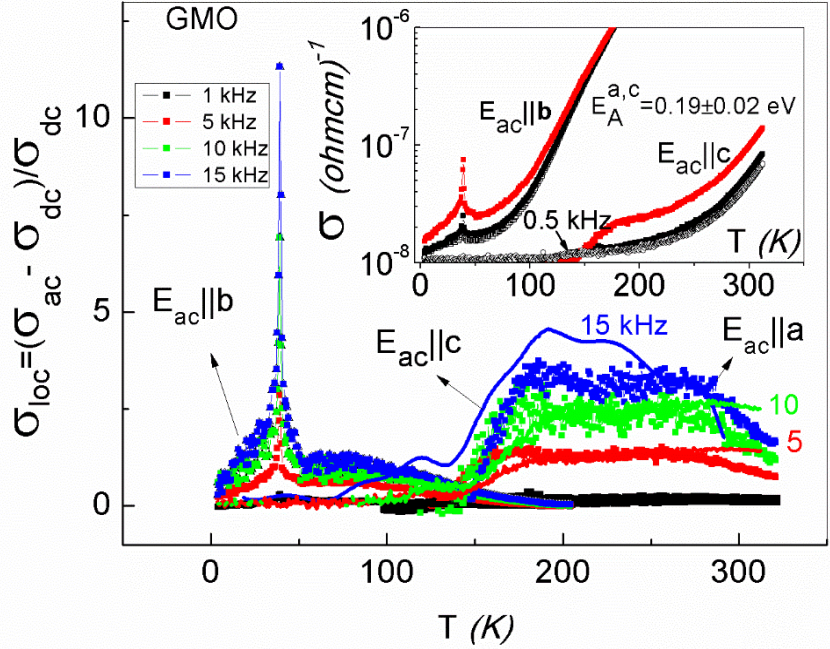

Fig. 3. Temperature dependences of local conductivity $\sigma_{loc} = (\sigma_{ac} - \sigma_{dc})/\sigma_{dc}$ for GMO measured along all the crystal axes at the frequencies indicated in the plots. Conductivity $\sigma_{loc}$ along the axis c is indicated by lines and along the axes a and b are indicated by dots. The inset shows temperature dependences of conductivity $\sigma$ measured along the axes a and c at the frequencies indicated in the plots.

The relative local conductivity $\sigma_{loc} = (\sigma_{ac}-\sigma_{dc})/\sigma_{dc}$ characterizes the ratio of the local conductivity to the percolation conductivity. Figure 3 shows the temperature ranges, along various axes of GMO, in which $\sigma_{loc}$ exceeds the percolation conductivity. The insert in Fig. 3 shows the temperature dependences of $\sigma$ along the axes b and c. Along axis b near $T_C$, there are also conductivity maxima.

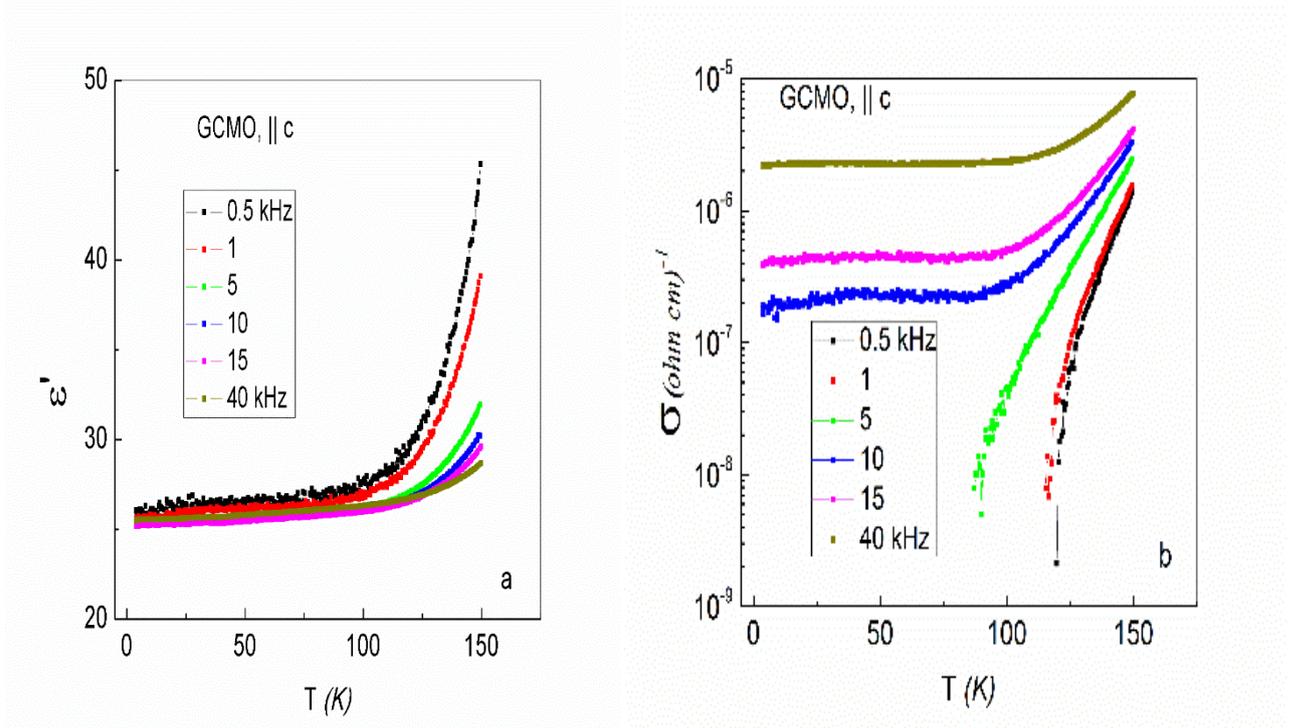

Fig. 4. Temperature dependences of (a) dielectric permittivity $\varepsilon'$ and (b) conductivity $\sigma$ for GCMO measured along axis c at the frequencies indicated in the plots.

It is seen, the percolation conductivity along the axis b begins to increase and exceeds $\sigma_{loc}$ near 100 K. Conversely, the local conductivity along the axes a and c begins to be observed only higher than 100 K and exceeds the percolation conductivity up to room temperature, slightly varying in magnitude. In this case, behaviors of $\sigma_{loc}$ along axes a and c are similar. The activation barrier at the boundaries of the restricted phase separation domains was determined from the jumps of $\sigma$ in the temperature range 90–150 K (the insert in Fig. 3) and was 0.2 eV in GMO. Note that the study of similar properties in a number of $RMn_2O_5$ (R = Eu, Gd, and Bi) [7, 9, 10] showed that, in all the crystals, a marked increase in $\varepsilon'$ and $\sigma$ begins near 100–125 K, and the activation barriers corresponding to this temperature dependence are close to be 0.2–0.3 eV. We suppose that they are due to a change in the states of the $Mn^{3+}$ and $Mn^{4+}$ ions. It turned out that similar situation took place also in GCMO [8]. Figures 4a and 4b show, for example, the temperature dependences of $\varepsilon'$ and $\sigma$ for GCMO along axis c at some frequencies; it is seen that $\varepsilon'$ and $\sigma$ increase at T> 100 K. In this case, it is seen that these values in GCMO significantly higher than those in GMO. These facts show that the phase separation domains in GMO and GCMO are similar, but their concentration in GCMO is significantly higher. Note that a new formation of restricted domains in $RMn_2O_5$ (R = Eu, Gd, and Bi) [7, 10] and doped $EuCeMn_2O_5$ and GCMO occurs [7, 8]. It is natural that, in this case, the local conductivity increases due to localization of carriers in deeper potential wells in a lattice with barriers higher than 0.5 eV. These barriers are also dependent on the type of R ions.

## 2.2. ELECTRIC POLARIZATION MEASURED BY THE PYROCURRENT METHOD

Figs 5a and 5b show the temperature dependences of the pyrocurrent and the polarization $P_b$ in GMO along the axis b, respectively. They were measured both in the "primary" crystal without preliminary polarization in the electric field and after a long (40-45 min) application of the polarizing field E = ± 3 kV/cm while cooling the sample from 300 to 5 K. Fig. 5a shows the intense narrow peak of the pyrocurrent near $T_C$ = 30 K, the amplitude of which only decreased after applying fields E = ± 3 kV/cm. This confirms the exchange–striction nature of the polarization in GMO along the axis b provided by a charge ordering of the $Mn^{3+}$–$Mn^{4+}$ ion pairs along this axis. In these ion pairs, the ferromagnetic and antiferromagnetic orientations of their spins alternate. A substantial difference of the strong double exchange of the ferromagnetic pairs and the weak indirect antiferromagnetic exchange leads at $T \leq T_C$ to the exchange striction and the central symmetry breaking of the crystal along the axis b [3]. The strong internal electric field along the axis b has an inhomogeneous staggered field-type structure. The application of a weak homogeneous external field E can only slightly disturb this internal field and decrease the polarization (Fig. 5b). Figures 5a and 5b show that there is asymmetry in the values of the pyrocurrent and the polarization when applying an external field along the internal field or opposite to it. The insert in Fig. 5a shows the temperature dependences of the pyrocurrent in the temperature range 50–150 K. Near 100 K, there are weak wide pyrocurrent maxima that are only observed in the case of preliminary polarization of the sample in fields E= ±3 kV/cm. This demonstrates the appearance of additional (significantly weaker) polarization that is also directed along the axis b and that exists in the temperature range 5–135 K. We assume that this polarization is due to restricted polar phase separation domains. From comparison of Fig. 5b and Fig. 3 it is seen that the correlation between temperature dependences of the local conductivity and the high-temperature polarization of GMO along the axis b takes place. Near 135 K, the free-dispersion $\sigma_{ac}$ of the phase separation domains is compared with the percolation conductivity ($\sigma_{loc} \approx 0$), and both the local conductivity and polarization $P_b$ have tails expanded in temperature.

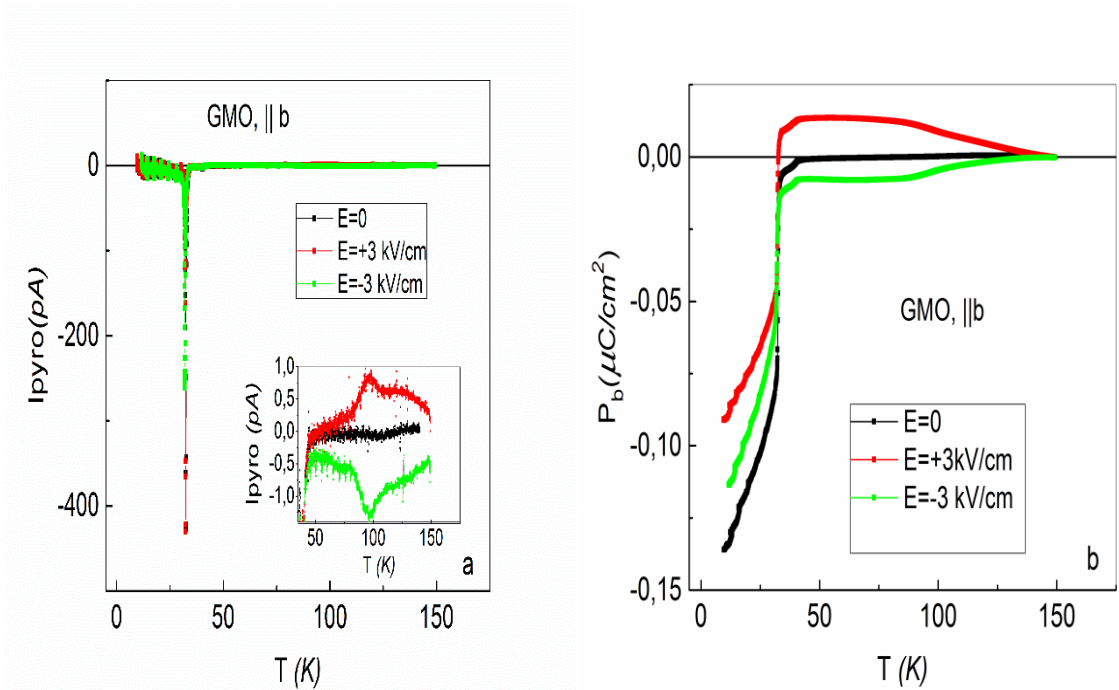

Fig. 5. Temperature dependences of (a) thermoactivated pyrocurrent Ipyro and (b) polarization Pb for GMO measured along the axis b in electric fields E= 0, ±3 kV/cm. The inset in (a) shows the temperature dependences of pyrocurrent Ipyro in an enlarged scale in the temperature range 50–150 K.

The temperature dependences of the pyrocurrent and the polarization for GCMO along the axis b are shown in Figs. 6a and 6b, respectively. Near $T_C$, the free-dispersion wide maximum of the pyrocurrent is observed at E= 0 with the intensity 250 time lower than that in GMO. This means that the initial charge ordering of ions $Mn^{3+}$ and $Mn^{4+}$ along the axis b is strongly disturbed by doping. The existence of this maximum demonstrates the conservation of the significantly weakened polarization of the exchange-striction nature in GCMO that is likely to be referred to the initial matrix. There is also polarization due to the polar phase separation domains that is observed only upon applying external field E= ±2.15 kV/cm. It is the limiting field that can be applied to the sample.

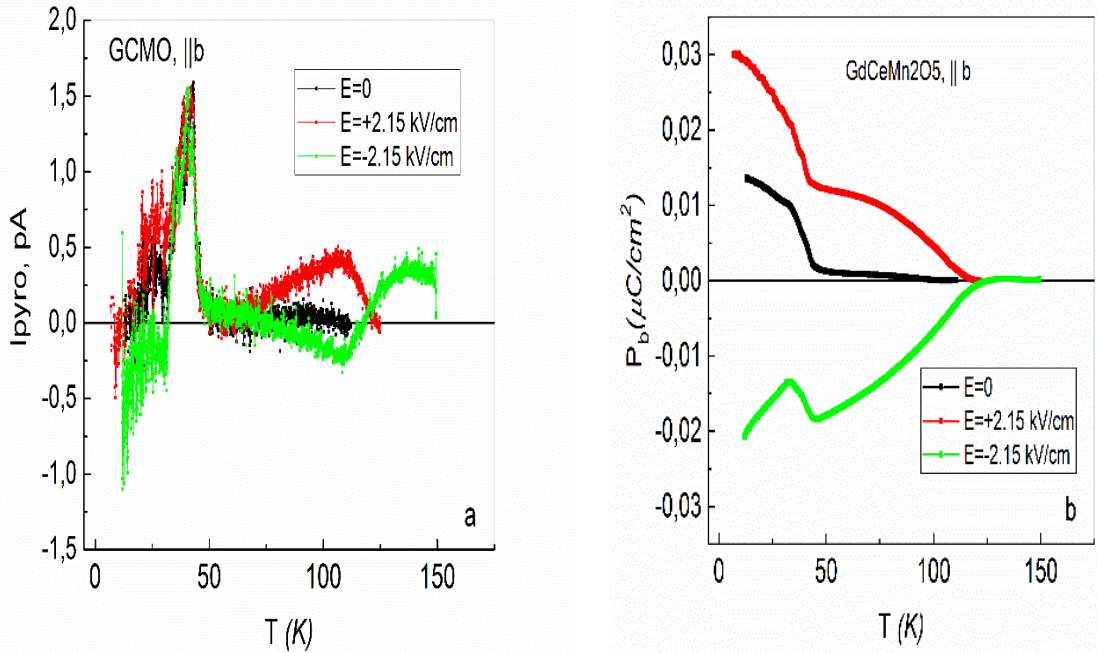

Fig. 6. Temperature dependences of (a) pyrocurrent Ipyro and (b) polarization Pb for GCMO measured along the axis b in electric fields E= 0, ±2.15 kV/cm.

In a higher field, the electric breakdown appears in the crystal. Note that an asymmetry in the values of the high-temperature polarization is observed for variously directed fields E.

The temperature dependences of the pyrocurrent and the polarization for GCMO along axis c are shown in Figs. 7a and 7b, respectively. In this direction, there is no low-temperature polarization of the exchange-striction nature, but the response of the restricted polar phase separation domains to electric field E||c as the pyrocurrent and the polarization exists. Despite that in GCMO along the axis c the dielectric permittivity and the conductivity are higher (Figs. 4a, 4b), we succeeded in applying higher field E= ±7.3 kV/cm. We relate this fact with a layer-by-layer distribution of the $Mn^{3+}$ and $Mn^{4+}$ ions in planes perpendicular to the axis c. Because of higher concentration of the phase separation domains in GCMO as compared to that in GMO, the 2D superstructure forms in GCMO [8]. This significantly decreases the percolation conductivity along the axis c and makes it possible to apply a higher field E. The observed values of the pyrocurrent and the polarization are twice as high as those that are observed in GCMO along the axis b. They exist in the same temperature range 5–135 K and are asymmetric with respect to variously directed

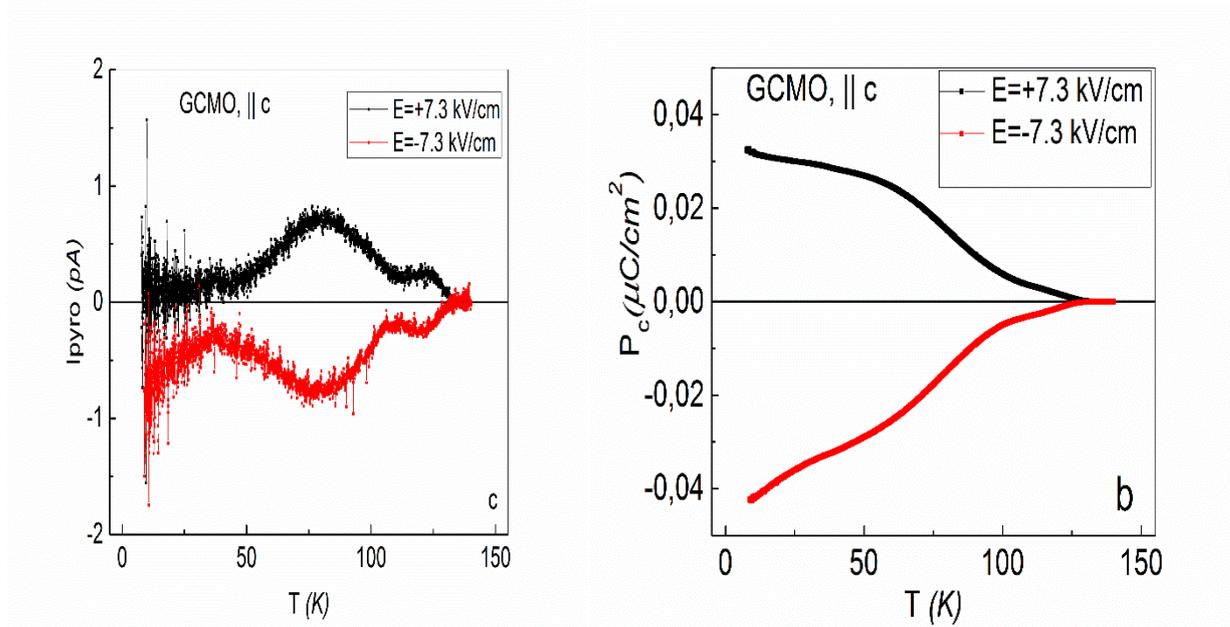

Fig. 7. Temperature dependences of (a) pyrocurrent Ipyro and (b) polarization Pc for GCMO measured along the axis c in electric fields E= 0, ±7.3 kV/cm.

fields ±E as well.

2.3. POLARIZATION MEASURED BY THE *PUND* METHOD. COMPARISON WITH THE PYROCURRENT METHOD

In this section, we perform a comparative analysis of the electric polarization measured by methods of pyrocurrent and hysteresis loops. When the hysteresis loops were measured by the PUND method, the *dynamic* hysteresis loops are studied as an immediate response of as-grown sample to short pulses of electric field. In this case, we measured the internal polarization, and the contribution of the conductivity to the polarization was avoided [9, 10]. In the case when the polarization is measured by the pyrocurrent method, the static polarization established, as noted above, as a result of long-term action of polarizing field E.

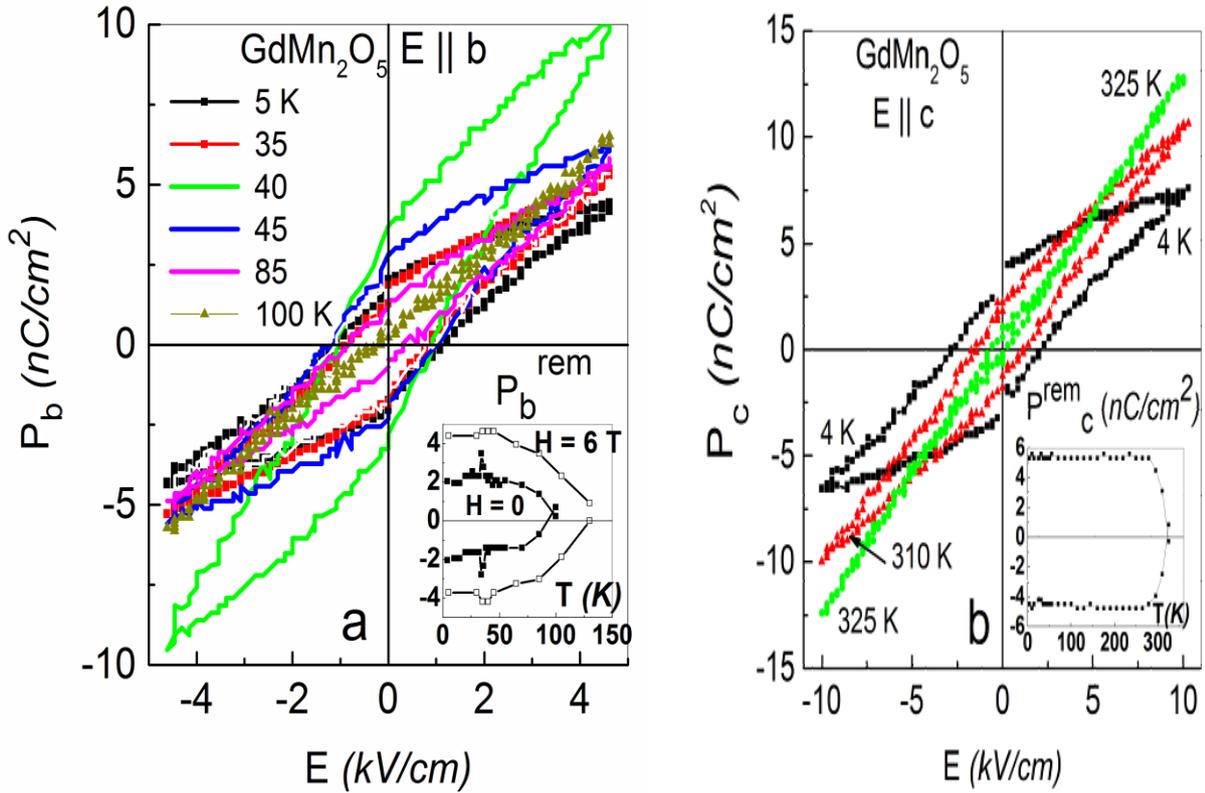

Fig. 8. Polarization hysteresis loops of GMO along axes b (a) and c (b) for some temperatures indicated in the plots. The insets show the temperature dependences of the remanent polarization. The inset in (a) shows the temperature dependences of the remanent polarization in applied magnetic fields H= 0 and 6 T.

Figure 8a shows the hysteresis loops of GMO along the axis b measured when applying pulses of electric field E, with duration of 2 ms and the amplitude to 5 kV/cm. We also studied the influence of applied magnetic field H= 6 T, H||b. A comparison of Fig. 8a and Fig. 5b shows that the polarization measured by pyrocurrent at E= 0 at T ≤ $T_C$ induced by the internal exchange-striction field is higher by a factor of 65 than the high-temperature polarization measured by the hysteresis loop method. Note that the remanent polarization of the hysteresis loops is only less by a factor of 6.5 than the high-temperature polarization provided by the restricted domains measured by the pyro-current method (insets in Fig. 8a and Fig. 5b). As is seen from the inset in Fig. 8a, the background residual polarization of the hysteresis loops is not dependent on temperature and near 100 K begins quite quickly decreases to zero. Near T= $T_C$= 30 K, the maxima of the remanent polarization are observed against this background. Because the low-temperature exchange-striction polarization forms in a strong internal field, it practically does not give the response in the hysteresis loops to a significantly weaker external field E. On the other hand, near $T_C$, when the internal field decreases sharply and fluctuations increase, the maxima of this remanent polarization is also observed in the loops.

Thus, the measurement of the GMO polarization along the axis b by two different methods enables us to separate the contributions of the exchange-striction polarization of the GMO matrix and the polarization due to polar phase separation domains. The first polarization is measured by the pyrocurrent method at E= 0; the second polarization is characterized by the remanent polarization of the hysteresis loops measured by the PUND method.

Since, at $T \leq T_C$, the pyrocurrent method in a polarizing field ±E measures the summary polarization of the matrix and the phase separation domains, in this case, an asymmetry of two the polarizations in the dependence on the sign of applied field E is observed (Fig. 5b). These polarizations are summed at the same orientation of the exchange-striction polarization of the matrix in E= 0 and the phase separation domains at E= −3 kV/cm. These polarizations are subtracted during the measurement in field E=+3 kV/cm. In the pyrocurrent method, the asymmetry of the polarizations due to the polar phase separation domains is observed in fields E= ±3 kV/cm at T> $T_C$, because of the long-term aftereffect of the induced polarization. However, in this case, the asymmetry is related to the sequencing of applying polarizing fields in subsequent measurement cycles. In Fig. 5b, field E= +3 kV/cm was applied in the first cycle.

As is seen from the inset in Fig. 8a, the application of magnetic field H||b, H= 6 T increases the remanent GMO polarization and also temperature, to which it exists. This indicates the magnetic nature of the remanent polarization and confirms the fact that the polarization is formed by the restricted phase separation domains. Actually, the double exchange is the main interaction that forms these domains in $RMn_2O_5$ [7,10, 27]. The double exchange increases the volume of

the phase separation domains and the barrier at their boundaries [6, 8, 12], which increases the polarization and increases the temperature at which $\sigma_{loc} \approx 0$.

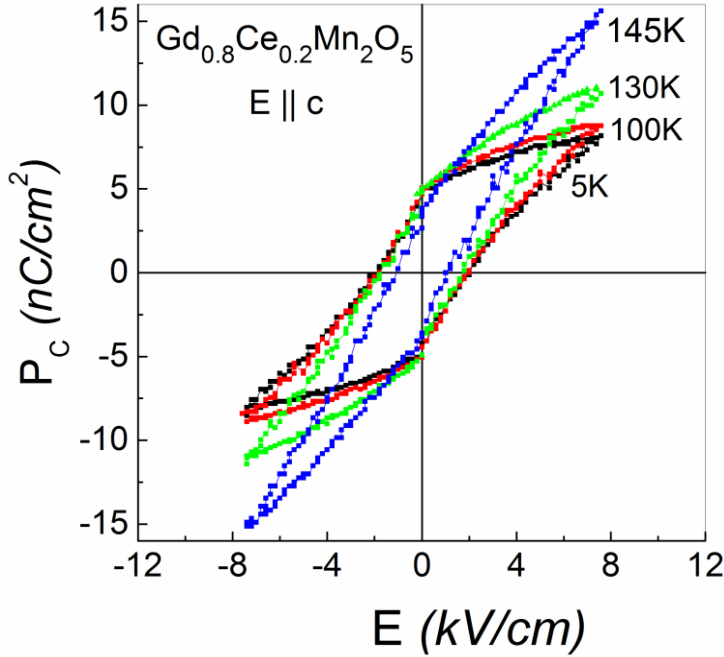

Fig. 9. Polarization hysteresis loops of GCMO along axis c for some temperatures indicated in the plots.

In GCMO, the exchange-striction polarization along the axis b is significantly weakened (Fig. 6b) and the conductivity (Fig. 2b) is quite high because of alternating the $Mn^{3+}$ –$Mn^{4+}$ ions along this axis, which does not allow the application of marked field E (E ≤ 2.15 kV/cm), because of the electrical breakdown of the sample. This is due to that the field E∥b increases the probability of the $e_g$ electron jumps between neighboring $Mn^{3+}$–$Mn^{4+}$ ion pairs and in the crystal matrix.

The inset in Fig. 8b shows the temperature dependence of the remanent polarizations of the hysteresis loops in GMO along the axis c. This polarization is not dependent on the temperature in a wide temperature range 5–295 K that coincides with the temperature range in which $\sigma_{loc}$ ∥c is not dependent on temperature and is higher than the percolation conductivity (Fig. 3). The remanent polarization disappears near T= 325 K, at which $\sigma_{loc} \approx 0$.

Figure 9 shows the temperature dependence of the GCMO hysteresis loops along the axis c. The remanent polarization is higher than that in GMO along the axis b by a factor of three, but it disappeared near T≈ 135 K, at which the conductivity dispersion begins to decrease sharply (Fig. 4b); i.e., $\sigma_{loc}$ tends to zero. Figure 9 shows, for example, the hysteresis loop at T= 145 K having the shape characteristic of a nonlinear dielectric. The main contribution in it is given by the conductivity. It seems likely that, in GCMO, the hopping conductivity between the phase separation domains increases due to a higher concentration of such domains, which increases the percolation conductivity at significantly lower temperature as compared to that of GMO. Thus, the electric polarization along the axis c induced by the restricted polar phase separation domains has a lower value, but it exists in GMO to a higher temperature than in GCMO.

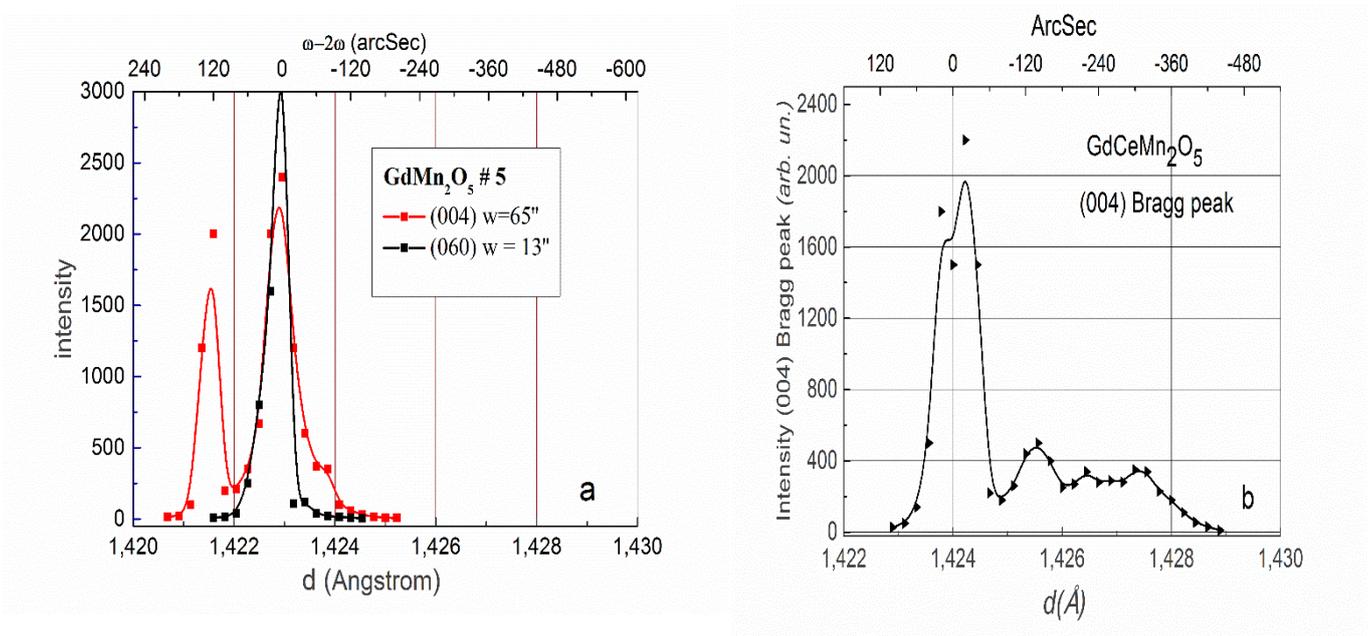

Fig. 10. Distribution of the intensities of the main Bragg reflections indicated in the plots in (a) GMO and (b) GCMO. The lower axis shows the interplanar spacings in corresponding directions, in angstrems. The upper axis shows the crystal rotation arc in the transverse plane to the chosen reflection direction, in angular seconds.

## 3. DISCUSSION

The exchange-striction nature of the low-temperature polarization along the axis b in $RMn_2O_5$ at $T \leq T_C$ has been revealed before [3]. We showed that this polarization in as-grown GMO and GCMO single crystals that had perfect crystal structures according to the X-ray diffraction data gave the maximum response as pyro-current and a homogeneous (single-domain) polarization in the internal field (at E= 0). The electric polarization of different nature was revealed in these crystals that existed from the lowest temperatures to some temperatures in the paramagnetic region, depending on the crystal type and for different axes. This polarization measured by the pyrocurrent method was significantly higher than the remanent polarization measured in the hysteresis loops by the PUND method.

Now we will discuss the properties and the nature of these revealed polarizations. The high-temperature polarization in GMO and GCMO is observed by both the measurement methods along different crystal axes, and its values and its disappearance temperatures are different along these axes as well. Thus, this polarization is not a result of a ferroelectric ordering. Actually, the temperature dependences of the dielectric permittivity and the conductivity do not contain free-dispersion maxima higher temperature $T_C$= 30 K.

As noted above, we relate the high-temperature polarization to the restricted polar phase separation domains. The phase separation is typical of manganites, including $RMn_2O_5$, containing ions $Mn^{3+}$ and $Mn^{4+}$ [6, 7, 12]. It is due to a finite probability of tunneling of $e_g$ electrons of the $Mn^{3+}$ ions to the neighboring $Mn^{4+}$ ions (double exchange [5, 6]). The phase separation and the self-organization of the charge carriers ($e_g$ electrons that recharge $Mn^{3+}$ and $Mn^{4+}$ ion pairs) are energetically preferable processes and lead to the formation of dynamically equilibrated phase separation domains at the balance of strong interactions. The double exchange and Jahn–Teller effect accumulate charge carriers ($e_g$ electrons) into the phase separation domains. Their Coulomb repulsion restricts the sizes of such domains [6, 7, 12]. Thanks to the balance of strong interactions, the formation of the dynamically equilibrated phase separation domains determines a number of their properties. First, the phase separation domains exist to rather high temperatures. Second, their response to external electric and magnetic fields is dependent on the values and duration of applied fields. The response of the dynamically equilibrated domains to weak short-time actions must fast relax to the initial state. It is important to take this fact into account as the polarization is measured by the PUND-method of hysteresis loops [9,10].

In order for the restricted domains to be polar and to have a local ferroelectric ordering, the fulfillment of the following conditions is necessary: they must have sizes that allow this ordering; their local symmetry must be noncentrosymmetric. Thus, the restricted polar domains must be structural ordered differently as compared to the centosymmetric Pbam matrix of crystals, and they must be matched to the initial matrix. To detect such polar domains, we studied the fine structure of the main Bragg reflections in GMO and GCMO using the high-resolution three-crystal X-ray diffractometer described in [7]. Figures 10a and 10b show the splitting of the Bragg peaks at room temperature in GMO and GCMO, respectively [8, 9]. The Bragg peak splittings are observed most clearly along the axis c (reflections (004)). GMO demonstrates the intense non-split Bragg reflection (060) corresponding to the initial space group Pbam. On the other hand, reflection (004) is split into two reflections with commensurable intensities. The more intense peak corresponds to the initial matrix with symmetry Pbam. Parameter c of the split-out peak is only changed in the third decimal place. Thus, there are two Bragg peaks belonging to two different but very close structures. We assume that the slightly distorted structure belongs to the polar phase separation domains. Note that the widths of both the Bragg peaks are small and characterize well-formed structures.

In GCMO, the situation is slightly different (Fig. 10b). We see two weakly split main Bragg peaks with close intensities and periodically repeated peaks characteristic of 2D superstructures. As a result, in GCMO at room temperature, a 2D superstructure consisting of the initial matrix layers and the phase separation domains forms along the axis c. As mentioned, this is due to a higher concentration of the phase separation domains as compared to that in GMO. In the case of GMO (Fig. 10a), we do not see periodic beats in the distribution of the intensities of the Bragg peaks. Because of this, we suppose that there are restricted phase separation domains in GMO at room temperature.

Because the phase separation domains spontaneously form in the initial matrix with symmetry Pbam due to the balance of strong internal interactions in this matrix (self-organization process) [6, 7, 12], these domains are similar and matched with the initial crystal matrix.

The polarity of the phase separation domains in $RMn_2O_5$ is determined by the two following factors. Inside the phase separation domains, the double exchange related to the transfer of $e_g$ electrons between the $Mn^{3+}$–$Mn^{4+}$ ion pairs lead to that the $Mn^{4+}$ ion position (oxygen octahedra) are occupied with Jahn–Teller $Mn^{3+}$ ions that deform these octahedra. The $Mn^{4+}$ ions with smaller size, in turn, occupy positions in noncentral pentagonal pyramids, and also distort it additionally. Both these factors lead to the non-centosymmetry of the phase separation domains and to their polarity. Note that the polarity of the phase separation domains forms spontaneously in the initial matrix and is not induced by external electric field. Thus, the restricted polar phase separation domains form the superparaelectric state in the initial matrices of GMO and GCMO.

Previously, the superparaelectric state of individual nanoscopic ferroelectric spheroidal regions in the centrosymmetric dielectric matrix was studied theoretically [16]. This state was experimentally studied for the first time in $RMn_2O_5$(R = Gd, Bi) [9, 10]. We assume that the polar phase separation domains are noncentrosymmetric and their sizes are sufficient for the ferroelectric single-domain state to appear in them. They are similar to the ferroelectric balls arranged in the centrosymmetric dielectric matrix that were studied in [16]. In this work, it was shown that the homogeneous polarization may arise in these particles–balls (in our case, in the phase separation domains) at low temperatures as their sizes R are smaller than the correlation radius Rc of the interaction between dipoles, but larger than the critical radius Rcr allowing the appearance of the ferroelectric order inside the domain. As these conditions are fulfilled, all dipoles are aligned in parallel in the polar domains. The surface screening of the depolarization fields makes the formation of the single-domain state of the restricted polar domains beneficial. If

R< Rcr, the individual paraelectric dipoles are not correlated and are restricted polar defects. In our case, they can only widen the initial Bragg peaks with the Pbam symmetry. The fact that GMO and GCMO clearly demonstrate two split well-determined Bragg reflections belonging to the phase separation domains and also to the initial matrices of the crystals shows that the conditions of the existence of the ferroelectric single-domain restricted regions presented in [16] are fulfilled.

In [16], it was also shown that the frozen superparaelectric state can form in assembles of ferroelectric nanoparticles in a dielectric matrix. In this state, the hysteresis loops and the remanent polarization as the responses to applied field E form at temperatures lower than the freezing temperature $T_f$. Temperature $T_f$ is determined from the condition that the potential barriers of reorientation of electric dipoles inside individual balls become equal to thermoactivation energies $\approx kT_B$. At $T> T_f$, the frozen superparaelectric is transformed to a usual superparaelectric, and the hysteresis loops disappear. In our case, temperatures $T_f$ can be referred to the high-temperature disappearance of the pyrocurrent in their maxima and to the disappearance temperature of the remanent polarization of the hysteresis loops as well. It was also taken in [16] that the temperature of the thermal disturbance of the ferroelectricity of individual balls must be much higher than $T_f$. This condition is also fulfilled in our crystals under study, since the polar phase separation domains are formed by strong interactions and at $T > T_f$ continue to exist representing the usual superparaelectric state.

Now, we discuss what information on the electric polarization of the restricted polar phase separation domains is obtained when studying this polarization by the two above mentioned methods. First, we note that the application of field E leads to the orientation of dipoles of individual phase separation domains along the field. In addition, field E can also transform the distribution of the $Mn^{3+}$ and $Mn^{4+}$ ions within the restricted domains, changing, in this case, their polarization. They are two processes that are observed to a different degree when measured the polarization by different methods. As the hysteresis loops are measured by the PUND method, when a field is applied as short-time pulses, the orientation of the dipoles of the individual restricted domains along the field appears predominantly and, to a lower degree, these domains are transformed. In this case, the duration of the electric field pulses and the intervals between them are chosen so that the summary response of the polarization of the polar domains and their conductivity in the first pulse was irreversible due to the polarization, whereas the response to the second pulse, to which the conductivity makes a predominant contribution, were closed [9,10]. This enables us to measure a partially relaxed internal initial polarization of the polar domains and to completely exclude the contribution of the conductivity to this polarization.

All three polarization sources are turned on when the polarization is measured by the pyrocurrent method. It seems likely that the induction of additional polarization along the field makes a predominant contribution as the polarizing field is applied for a long time in the wide temperature range 5–330 K. Actually, the application of field E along a fixed axis initiates the drift of valence $e_g$ electrons inside the restricted phase separation domains along this axis. These electrons recharge ions $Mn^{3+}$ and $Mn^{4+}$ within the domains. As a result, the distribution of ions $Mn^{3+}$ and $Mn^{4+}$ and the spatial distribution of local distortions are changed inducing the polarization along field E. The inertia of such a change in the states of the phase separation domains makes it possible to obtain the maximum pyrocurrent during subsequent heating of the sample in field E= 0. In addition to this contribution, there are naturally contributions of the orientation of these transformed restricted polar domains and the conductivity. Because of this, the polarization obtained by the pyrocurrent method is significantly higher than that measured by the hysteresis loop method and not only due to the contribution of the conductivity. In this case, the remanent polarization of the hysteresis loops gives the most correct information on the internal polarization due to the displacements of ions in noncentral phase separation domains and that spontaneously appears in the initial crystal matrix at E= 0.

Note that there is an opposite action of applied electric E and magnetic H fields on the polarization of the restricted polar domains. Field E, inclining the barriers at domain boundaries, increases the leakage of a part of $e_g$ electrons from these domains and from the whole sample through the external circuit of the capacitor. On the other hand, the magnetic field increases the double exchange and thus increases the barriers at the restricted domain boundaries. In this case, the leakage of $e_g$ electrons that is determined by $kT_B$ decreases. As noted above, it is precisely the number and the distribution of the $e_g$ electrons that recharge $Mn^{3+}$ and $Mn^{4+}$ ion pairs within the restricted polar domains determine the distribution of these ions within the domains, immediately influencing their polarization. A long-term application of an electric field to the phase separation domains (in the pyrocurrent method) changes their polarization up to the temperature at which the kinetic energy of percolation electrons becomes equal to the barrier height at the restricted polar domain boundaries. At the same temperature, the condition $\sigma_{loc} \approx 0$ is fulfilled. The pyrocurrent maximum is observed near this temperature. Because the conditions on the temperatures at which the remanent polarization in the hysteresis loops disappears are the same ($\sigma_{loc} \approx 0$), the polarizations of various values measured along different axes disappear near the same temperatures in both these methods. Both the factors (the pyrocurrent maximum and the disappearance of the remanent polarization in the hysteresis loops) simulate the behavior near the ferroelectric phase transition. Actually, at these temperatures, the frozen superparaelectric state of the restricted polar domains is transformed to the usual superparaelectric state. Thus, the pyrocurrent maxima and the disappearance of the remanent polarization in the hysteresis loops are the necessary but not sufficient conditions of the ferroelectric phase transition. The restricted charge domains or the frozen superparaelectric state of the dielectric restricted polar domains can give similar response to an applied electric field.

## 4. CONCLUSIONS

It has been established that, in perfect $GdMn_2O_5$ and $Gd_{0.8}Ce_{0.2}Mn_2O_5$ single crystals, the homogeneous polarization forms in the ferroelectric state in the internal exchange-striction field of the staggered field type at $T \leq T_C = 30$ K. The external electric field $E \neq 0$ only decreases this polarization. The existence of manganese ions with different valences ($Mn^{3+}$ and $Mn^{4+}$) in $GdMn_2O_5$ and $Gd_{0.8}Ce_{0.2}Mn_2O_5$ leads to spontaneous formation of the restricted polar phase separation domains. These domains are responsible for the temperature evolution of the dielectric permittivity and the local conductivity. The restricted polar domains form the superparaelectric state in $GdMn_2O_5$ and $Gd_{0.8}Ce_{0.2}Mn_2O_5$. The domain sizes and the correlation radius of interaction of the electric dipoles are such that the frozen superparaelectric state is established in them at $T \leq T_f$, $T_f \gg T_C$. In this state, the response to applied field E has a shape of hysteresis loops with the remanent polarization. Temperature $T_f$ is determined by the condition of the equality of the kinetic energy of charge carriers to the barrier heights at the restricted polar domain boundaries. The existence of the restricted polar phase separation domains was confirmed by the splitting of the Bragg reflections (004) into two narrows peaks. These reflections belong to two well-formed structural formations that belong to the crystal matrix and the restricted polar domains. Lattice parameters c in these structural formations are different in the third decimal place.